\documentclass[twocolumn,showpacs,preprintnumbers,amsmath,amssymb]{revtex4}
\usepackage{mathbbold}
\usepackage{multirow}

\usepackage{graphicx}
\usepackage{dcolumn}
\usepackage{bm}

\begin{document}


\title{Promotional effect on cold start problem and diversity in a data characteristic based recommendation method}

\author{Tian Qiu $^{1}$\footnote{email: tianqiu.edu@gmail.com}}
\author{Zi-Ke Zhang $^{2,3,4}$\footnote{email: zhangzike@gmail.com}}
\author{Guang Chen $^{1}$}
\affiliation{$^{1}$ School of Information Engineering, Nanchang Hangkong
University, Nanchang, 330063, P.R. China \\$^{2}$ Institute of Information Economy, Hangzhou Normal University - Hangzhou 310036, P. R. China \\$^{3}$ Web Sciences Center, University of Electronic Science and Technology of China - Chengdu 610054, P.R. China \\$^{4}$ Beijing Computational Science Research Center, Beijing 100084, P. R. China}


\date{\today}

\begin{abstract}
Pure methods generally perform excellently in either recommendation accuracy or diversity, whereas hybrid methods generally outperform pure cases in both recommendation accuracy and diversity, but encounter the dilemma of optimal hybridization parameter selection for different recommendation focuses. In this article, based on a user-item bipartite network, we propose a data characteristic based algorithm, by relating the hybridization parameter to the data characteristic. Different from previous hybrid methods, the present algorithm adaptively assign the optimal parameter specifically for each individual items according to the correlation between the algorithm and the item degrees. Compared with a highly accurate pure method, and a hybrid method which is outstanding in both the recommendation accuracy and the diversity, our method shows a remarkably promotional effect on the long-standing challenging problem of the cold start, as well as the recommendation diversity, while simultaneously keeps a high overall recommendation accuracy. Even compared with an improved hybrid method which is highly efficient on the cold start problem, the proposed method not only further improves the recommendation accuracy of the cold items, but also enhances the recommendation diversity. Our work might provide a promising way to better solving the personal recommendation from the perspective of relating algorithms with dataset properties.
\end{abstract}


\maketitle

\section{INTRODUCTION}

Favored by increasing information, people can enjoy an abundant life, however, people are also brought into a quandary decision of getting what they
actually prefer. For example, how to select a satisfactory dress from various dress brands, or get an interesting book to read from
the book sea \cite{mas04}. As a powerful tool, recommendation engine emerges to help people out of the overloaded information \cite{ado05}. With an inquiry of personal recommendation, developing efficient recommendation methods has become one of the central scientific programs.

A great many algorithms have been proposed, and have led to a considerable progress, such as the collaborative filtering (CF) algorithm \cite{gol92,sch07}, the content based algorithms \cite{paz07}, and the relevant extensive studies \cite{bal97,gol01,mas01,hof04,ble03,lau06,ren08,zha11}. Recently, favored by the fruitful achievements of complexity theory, complex network based recommendation algorithms have been proposed, which directs a promising way for the personal recommendation \cite{zha07,zho07,jia08,zho09,liu09a,zho10,liu09b,mar08,zha10a,sha10,liu09c,liu10a}. Meanwhile, concepts from traditional physical domain have been introduced into the algorithm design, e.g., the thought of mass diffusion \cite{zho07,zho10} and heat conducting \cite{zha07,zho10}, which greatly promotes recommendation accuracy and diversity.

Most previous studies can be classified into two categories, i.e., the pure algorithms and the hybrid algorithms. The pure method refers to a single algorithm without any algorithm combination, such as the standard CF method\cite{gol92}, whereas the hybrid algorithm refers to the algorithm which combines different pure algorithms, such as the collaborative filtering and the content hybrid method \cite{bal97}, the heat conducting and the probability spreading hybrid method (HHP)\cite{zho10}, and their variants \cite{lv11, liu11}. Considering the well accepted evaluators of personalized recommendation, i.e., the recommendation accuracy and the diversity, the pure methods generally either perform excellently in the accuracy but show par performance in the diversity, or vice versa. For instance, the probability spreading (PBS) method \cite{zho07} shows a great advantage in recommendation accuracy but a less outstanding performance in diversity, whereas the heat conducting (HTS) method \cite{zha07} greatly improves the recommendation diversity but at the cost of the accuracy. Therefore, different hybrid methods have been proposed in order to improve the performance of both accuracy and diversity. The hybrid algorithm relates different pure methods via some function form, with the recommendation performance usually controlled by the value of the hybridization parameter. Generally speaking, the hybrid method indeed performs better in the both aspects of the recommendation accuracy and the diversity than the pure method at some optimal hybridization parameter.

However, how to find out the optimal hybridization parameter still remains controversial. For most hybrid methods, the optimal hybridization parameters obtained from different evaluators are different. By far, most algorithms take a one-elevator optimal parameter selection, namely, choosing the optimal value according to the recommendation performance of one evaluator, e.g., the recommendation accuracy. However, without bias, different recommendation focus might prefer different evaluator performance. Consequently, a challenging question emerges: which evaluator should be taken as a common basis of optimization? Even though the recommendation accuracy is widely accepted to be the most important proxy in personalized recommendation, the cold start problem or the recommendation diversity also raises a central interest \cite{zha10b,zho10,ahn08}. The cold start problem refers to how to recommend the new item or recommend the interesting item to new users due to the lackness of activity records. The diversity and novelty also significantly mark the vitality of a system. Explicitly, one can hardly find out the same value of the optimal hybridization parameter according to different recommendation focal purposes. Moreover, even when evaluating the recommendation accuracy, different indicators might correspond to different optimal hybridization parameter value. For example, the ranking score \cite{zho07} and the precision \cite{her04} are both indicators which are used to evaluate the recommendation accuracy. However, the optimal parameters obtained by the ranking score and the precision are not usually consistent for a hybrid method \cite{lv11}.

Motivated by the explicit dilemma to choose a proper parameter for hybrid algorithms, in the present paper, we propose a data characteristic based algorithm (DCB) by finding out the possible correlation between the hybridization parameter and the data characteristic represented by item degrees. With this implementation, instead of using only one evaluator as the basis of optimal hybridization parameter selection, the optimal parameter is adaptively assigned for each specific individual item according to the correlation between the algorithm and the dataset property. By testing our algorithm on three datasets, our algorithm shows a great promotional effect on the cold start problem and the recommendation diversity, while simultaneously exhibits a high recommendation accuracy.

The remainder of this paper is organized as follows. In the next section, we detail the bipartite network and the investigated algorithms of the recommendation system. Some popular indicators to evaluate the recommendation algorithm performance are introduced in Section III. Then, we compare the results of the present algorithm with a highly accurate pure algorithm, a hybrid method with both high accuracy and high diversity, and even an improved hybrid method which is highly efficient on cold start problem. Finally comes to the conclusion.

\section{ALGORITHMS}

A recommendation system can be described by a bipartite network composed of a user set and an item set. The user set includes $m$ users $U =\{u_{1}, u_{2}, . . . u_{i}, . . ., u_{m}\}$, and the item set includes $n$ items  $O = \{ o_{1}, o_{2}, . . ., o_{\alpha}, . . ., o_{n}$. If an item $o_\alpha$ is collected by
a user $u_i$, then add a link between them. The adjacent
matrix which links the users and the items is $A=\{a_{i\alpha}\}$. If the item $o_\alpha$ is collected by the user $u_i$, then $a_{i\alpha} = 1$, otherwise, $a_{i\alpha} = 0$. The degree of an item is denoted as the number of links owned by the item. We assume an item is popular if the item has a big degree, otherwise, the item is cold. The task of a recommendation algorithm is to provide a user a ranking list of items that the user does not collect, and then recommend the items with higher rankings for the user.

In the following algorithms, a so-called ''resource'' is introduced to items. If we label the initial level of resource by a vector $f_{0}=[f_{1,0}^{i}, ... , f_{n,0}^{i}]$, the final resource of the item $f=[f_{1}^{i}, ... , f_{n}^{i}]$ is obtained according to a resource redistribution process described by a transformation form,

\begin{equation}
f=Wf_{0}, \label{eq.3}
\end{equation}
where $W$ is the resource reallocation matrix. By ranking the level of the final resources, the items with higher resources will be recommended to users. Therefore, how to redistribute the resources plays a key role in the recommendation process.

The mass-diffusion based algorithm, refering to the PBS, is reported as a highly accurate method. The PBS is actually a three-step random walk process. The item firstly distributes the resource to its neighboring users with an equal probability, while the user again redistribute its total level of resource to its neighboring items. The item then obtains its final level of resource by summing up all the resources from its neighboring users. The resource transformation matrix the PBS is as,

\begin{equation}
W_{\alpha\beta}^{H}=\frac{1}{k_{\alpha}}\sum
_{j=1}^{u}\frac{a_{\alpha j}a_{\beta j}}{k_{j}}, \label{eq.2}
\end{equation}
where $k_{\beta}$ is the degree of item $o_{\beta}$. The PBS achieves a high recommendation accuracy for assigning more resources on popular items, however, potentially puts the recommendation diversity at risk.

By incorporating heat-conducting analogous process, the HTS method is proposed with a similar random-walk resource redistribution process. Firstly, the user receives an average level resource from its neighboring items, and then the item again gets a feedback of the average resource from its neighboring users. The transformation matrix then reads,

\begin{equation}
W_{\alpha\beta}^{H}=\frac{1}{k_{\alpha}}\sum
_{j=1}^{u}\frac{a_{\alpha j}a_{\beta j}}{k_{j}}, \label{eq.2}
\end{equation}
where $k_{\alpha}$ is the degree of item $o_{\alpha}$, and $k_{j}$ is the degree of the user $u_{j}$. Differently from the PBS, the HTS assigns more resources on cold items, and therefore shows a good performance in recommendation diversity, but at the cost of the recommendation accuracy.

To achieve a high accuracy and diversity of recommendation, a hybrid method (HHP) is proposed \cite{zho10}, by elegantly combining the heat conduction and the mass-diffusion method as,

\begin{equation}
W_{\alpha\beta}^{H+P}=\frac{1}{k_{\alpha}^{1-\lambda}k_{\beta}^{\lambda}}\sum
_{j=1}^{u}\frac{a_{\alpha j}a_{\beta j}}{k_{j}}, \label{eq.4}
\end{equation}
where $\lambda \in [0,1]$. When tuning the hybridization parameter $\lambda$ to a suitable value, the HHP method outperforms in both the recommendation accuracy and the diversity.

Based on the HHP method, an improved hybrid method (OHHP) is proposed \cite{qiu11a}, focusing on resolving the cold-start problem. In the OHHP, individual item degrees are incorporated into the hybridization parameter formula of the original HHP, where the hybridization parameter reads,

\begin{equation}
\lambda=(\frac{k_{\beta}}{k_{max}})^{\gamma}, \label{eq.5}
\end{equation}

\begin{figure}[htb]
\centering
\includegraphics[width=6cm]{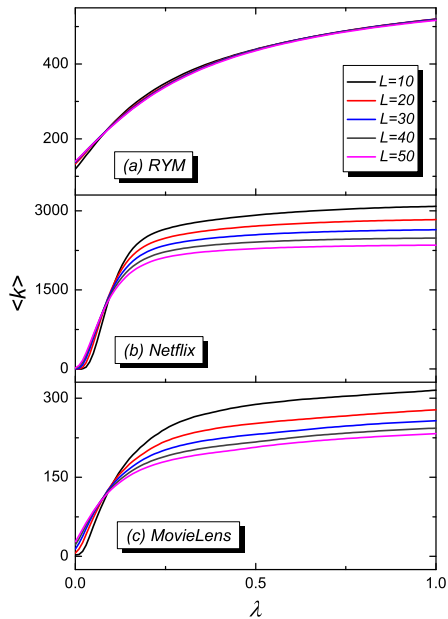}
\caption{The average degree $\langle k \rangle$ on the hybridization
parameter $\lambda$ is displayed. \label{Fig:1} }
\end{figure}
where $k_{\beta}$ is the degree of the examined object, $k_{max}$ is the maximum degree of all the objects, and $\gamma$ is a tuned parameter. The OHHP actually optimizes the probability spreading factor in the transformation matrix of Eq. (4) according to the individual item degree level, therefore it greatly enhances the recommendation accuracy of cold items, whereas keeps a high recommendation accuracy of the overall and the popular items.

Compared with pure methods of the PBS and the HTS, the hybrid methods show a great advantage in both the recommendation accuracy and diversity, however, they suffer from the hybridization parameter selection for different recommendation focal purposes. The explicit dilemma inspires us to find out an efficient hybridization parameter adaption procedure. In the OHHP, the cold start problem is well resolved by considering the item degree into the parameter selection procedure, which indicates a promising way to design the data characteristic based algorithms. The degree provides a simple way to describe the dataset property. If a universal relation can be revealed between the hybridization parameter and the average degree by some function form $\lambda \sim f(\langle k \rangle)$, the parameter $\lambda$ then can be adaptively assigned. However, generally, for different recommendation list length $L$, the relation function between $\lambda$ and the degree is different. As shown in Fig. 1, for all three datasets, i.e., the \emph{RYM}, the \emph{Netflix}, the \emph{MovieLens} (Details of the three datasets will be introduced in Section IV), the average degree of the items on the hybridization parameter $\lambda$ of the HHP exhibits a different behavior for different recommendation list length $L$, which suggests that one should not provide a uniform relational function between the average degree and the hybridization parameter $\lambda$.

\begin{figure}[htb]
\centering
\includegraphics[width=6cm]{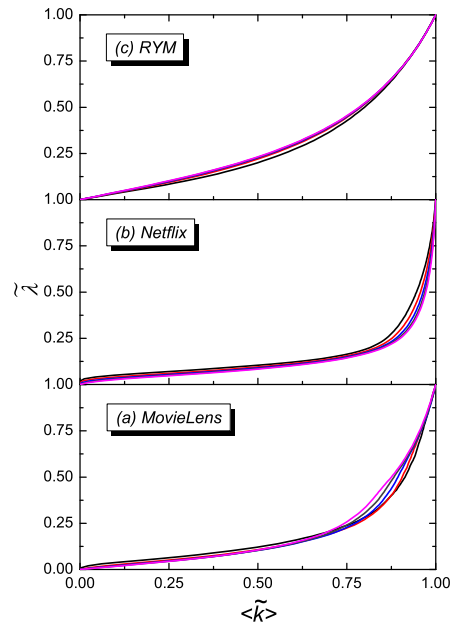}
\caption{The rescaled hybridization parameter $\widetilde{\lambda}$ vs. the rescaled average degree $\langle \widetilde{k} \rangle$ is displayed. \label{Fig:2} }
\end{figure}

In order to obtain a scaling behavior of the relation between $\langle k \rangle$ and  $\lambda$ independent of the recommendation list length $L$, we analytically investigate the \emph{recommendation bias} for the hybrid algorithm. On average, the probability that a target user $i$ collects an item $\beta$ is directly proportional to $\beta$'s degree, $k_{\beta}$, that is to say, $a_{i\beta} \propto \frac{k_{\beta}}{n}$, where $n$ is the number of items.
Based on the theoretical analysis in \cite{qiu11a}, we hypothesize that the probability of $a_{i\beta}$ is independent of other links, and then the expected score of each user-item link, $f_{i\alpha}$, can be calculated as

\begin{equation}
\begin{array}{rcl}
f_{i\alpha} &\propto& k_{\alpha}^{\lambda}\sum\limits_{\beta}k_{\beta}^{2-\lambda}
\\  &\propto& k_{\alpha}^{\lambda}\int f(k) k^{2-\lambda}dk,
\end{array}
\end{equation}
where $f(k)$ is the probability distribution function of the item degrees. As suggested in Ref. \cite{qiu11a}, $f(k)$ obeys a power-law distribution, i.e., $f(k)\propto k^{-\nu}$. Then, one can calculate $f_{i\alpha}$ as,

\begin{equation}
\begin{array}{rcl}
f_{i\alpha} &\propto& k_{\alpha}^{\lambda}\int k^{2-\lambda-\nu}dk
\\  &\propto&  k_{\alpha}^{\lambda} (k^{3 - \lambda- \nu} \mid _{k_{min}}^{k_{max}}),

\end{array}
\end{equation}
where $k_{max}$ and $k_{min}$ are respectively the maximum and the minimum of the item degrees, with $k_{max}\gg k_{min}$. Assuming $M=\frac{k_{min}}{k_{max}}$, one then obtains,

\begin{equation}
\begin{array}{rcl}
f_{i\alpha}  &\propto&  (\frac{k_{\alpha}}{k_{max}-k_{min}})^{\lambda}(1-M)k_{max}^{3-\nu}
\\  &\propto&  (\frac{k_{\alpha}}{k_{max}-k_{min}})^{\lambda},
\end{array}
\end{equation}

Instructed by the theoretical analysis in Eq. (8), in our study, we rescale $\lambda \sim f(\langle k \rangle)$ as $\widetilde{\lambda} \sim f(\langle \widetilde{k} \rangle)$ by normalizing the $\lambda$ dependent average degree $\langle k(\lambda) \rangle$ to $\langle \widetilde{k}(\widetilde{\lambda}) \rangle$ as,

\begin{equation}
\langle \widetilde{k} (\widetilde{\lambda}) \rangle=\frac{\langle k(\lambda) \rangle-k
_{min}}{k_{max}-k_{min}},
\end{equation}
where $\langle ... \rangle$ takes an average on all the items in the recommendation list, $k_{max}=max\{\langle k(\lambda) \rangle_{\beta \in L, \lambda \in [0,1]}\}$ and $k_{min}=min\{\langle k(\lambda) \rangle_{\beta \in L, \lambda \in [0,1]}\}$. The rescaled procedure assures $\langle \widetilde{k} \rangle \in[0,1]$. The $\widetilde{\lambda}$ on the rescaled average degree $\langle \widetilde{k} \rangle$ for the recommendation list length $L=10$, 20, 30, 40 and 50 is shown in Fig. 2, where a satisfactory data collapse independently from thee recommendation list length $L$ is observed for the \emph{RYM}, the \emph{Netflix} and the \emph{MovieLens}. We fit the rescaled $\widetilde{\lambda} \sim f(\langle \widetilde{k} \rangle)$ with a combined exponential function form which reads,

\begin{equation}
\widetilde{\lambda}=ae^{b \langle \widetilde{k}
\rangle}+ce^{d\langle \widetilde{k} \rangle},
\end{equation}

The corresponding coefficient is $(a, b, c, d)=(0.04,3.31,-0.04,-12.28)$ for the \emph{RYM}, $(a, b, c, d)=(0.03,2.25,1.75\times10^{-9},19.78)$ for the \emph{Netflix}, and $(a, b, c, d)=(0.03,2.48,4.95\times10^{-7},14.05)$ for the \emph{MovieLens}. For a specific individual item $\beta$, we replace the rescaled average degree $\langle \widetilde{k} \rangle$ in Eq. (10) with a normalized individual item degree,

\begin{equation}
\widetilde{k}_{\beta}=\frac{(k_{\beta}-k_{min})}{(k_{max}-k_{min})},
\end{equation}
where $k_{\beta}$ is the degree of the examined item $\beta$. The hybridization parameter of the item $\beta$ is then adaptively assigned by,

\begin{equation}
\widetilde{\lambda}=ae^{b \widetilde{k}_{\beta}
}+ce^{d\widetilde{k}_{\beta}},
\end{equation}

\section{METRICS}
Recommendation accuracy is with no doubt one of the most important indicators to evaluate the performance of an algorithm. As an adjunct to accuracy, recommendation diversity is addressed to be an important elevator to quantify the personal recommendation. In our study, we take the ranking score and the precision to quantify the recommendation accuracy, the inter-diversity and the inner-diversity to quantify the recommendation diversity. Moreover, to specifically investigate the recommendation accuracy of cold items, we further study an item-dependent ranking score and an item-dependent precision.

\subsection{RECOMMENDATION ACCURACY}


\textbf{RANKING SCORE ($r$)}\cite{zho07}\textbf{.-}The ranking score $r_{\alpha i}$ for the item $o_{\alpha}$ to the user $u_{i}$
is defined as,
\begin{equation}
r_{\alpha i}=\frac{p_{\alpha}}{n-k_{i}}.
\end{equation}
where $n$ is the number of all items, $k_{i}$ is the degree of the user $u_{i}$, and $p_{\alpha}$ is the
position of the recommended item $o_{\alpha}$ located in all the uncollected
items of the user $u_{i}$. Generally speaking, users collect the items which they prefer. Namely, for a user $u_{i}$,if the deleted link with an item $o_{\alpha}$ is in a higher rank of $u_{i}'s$ all deleted links, the algorithm is more accurate. The average ranking score $r$ is then defined as the average of $r_{\alpha i}$ over all the deleted links. The smaller the $r$, the more accurate the algorithm.

To focus on the recommendation accuracy of cold items, we define an item-degree dependent ranking score $r_{k}$ as the average ranking score over items with the same value of degrees\cite{zho08}.


\textbf{PRECSION ($P$)}\cite{her04}\textbf{.-}The recommendation precision $P$ is defined as

\begin{equation}
P=\frac{1}{m}\frac{\sum_{i}^{m}q_{iL}}{L}, \label{eq.7}
\end{equation}
where $q_{iL}$ is the number of the user $u_{i}'s$ deleted links contained in the top $L$ recommended item list. The larger the
$P$, the higher accuracy the algorithm.

Similarly, to better understand the recommendation accuracy of the cold items, we define an item-degree dependent precision by,

\begin{equation}
P_{k}=\frac{1}{m}\frac{\sum_{i}^{m}q_{iL}^{k}}{L}, \label{eq.11}
\end{equation}
where $q_{iL}^{k}$ is the number of the user $u_{i}'s$ deleted links for
items with degree $k$ in the top $L$ recommended item list.

\subsection{RECOMMENDATION DIVERSITY}


\textbf{INTER DIVERSITY ($D_{Inter}$).-}$D_{Inter}$ quantifies the difference between different users
recommendation list by

\begin{equation}
D_{Inter}=\frac{2}{m(m-1)}\sum_{i=1}^{m}\sum_{j=i+1}^{m}(1-\frac{\left({\L}_{i}
\bigcap {\L}_{j}\right)}{L}), \label{eq.8}
\end{equation}
where $\left({\L}_{i} \bigcap {\L}_{j}\right)$ is the number of common
recommended items for user $u_{i}$ and $u_{j}$ in the top $L$
recommendation list. Generally, the greater the $D_{Inter}$, the more personal the
recommendation for different users, and vice versa.

\textbf{INNER DIVERSITY ($D_{Inner}$).-}$D_{Inner}$ calculates
the difference within a specific user recommendation list by

\begin{equation}
D_{Inner}=\frac{1}{mL(L-1)}\sum_{i=1}^{m}\sum_{\alpha \neq
\beta}(1-S_{\alpha\beta}), \label{eq.9}
\end{equation}
where $S_{\alpha\beta}=\frac{1}{\sqrt{k_{\alpha}k_{\beta}}}\sum_{i=1}^{m}
a_{\alpha i}a_{\beta j}$ is the cosine similarity between items
$o_\alpha$ and $o_\beta$ in a single user's top $L$ recommended
item list. Generally, the greater the $D_{Inner}$, the higher
diversification of the recommendation list for a specific user, and
vice versa. Thus, a large $D_{Inner}$ provides an evidence
that the algorithm can potentially enlarge visions of each single
user by recommending less similar items.

\section{DATA}
We test the algorithm performance on three datasets, the \emph{RYM}, the \emph{Netfilx} and the \emph{MovieLens}. The \emph{RYM} is a music
rating system with a ten-level rating, and the \emph{Netflix} and the \emph{MovieLens} are movie rating systems with a five-level rating. The \emph{RYM} dataset is downloaded from the music rating web site RateYourMusic.com, the \emph{Netflix} dataset is obtained by randomly selecting from the huge dataset of the \emph{Netflix} Prize, and the \emph{MovieLens} is downloaded from the web site of GroupLens Research. Due to the different level of ratings, we perform a coarse-graining mapping to a unary form for all the three datasets. If the rating is no less than three for the \emph{Netflix} and the \emph{MovieLens}, and six for the \emph{RYM}, we argue that the item is collected by a user. The \emph{Netflix} contains 10000 users, 6000 items, and 701947 links, and the \emph{MovieLens} contains 943 users, 1682 items, and 82520 links. When dealing with the \emph{RYM}, several items are found to have particularly large degrees, which are much higher than the rest of items. We then remove about 10 items with a huge number of degrees, and the \emph{RYM} then contains 33786 users, 5381 items and 613387 links. The sparsity of the datasets, defined as the number of links proportional to the total number of the user-item links, is $3.37 \%$, $1.17 \%$, and $5.20 \%$ for the  \emph{RYM}, the \emph{Netflix}, and the \emph{MovieLens}, respectively.

We divide a dataset into two subsets of the training set and the test set. We randomly cut the $10\%$ links as the test set, and remain the rest $90\%$ links as the training set. We utilize the training set to make predictions for users, and the test set to test the algorithm performance.

\begin{table*}
\caption{The overall ranking score $r$, the item-degree dependent ranking score
$r_{k\leq10}$, the overall precision $P$, the item-degree dependent precision
$P_{k\leq10}$, the inter-diversity $D_{inter}$, and the inner-diversity $D_{inner}$ of the PBS, the HHP, the OHHP, and the DCB methods are shown for the \emph{RYM},
the \emph{Netflix}, and the \emph{MovieLens}, with $L=50$.}
\begin{center}
\begin{tabular}{cccccccc}
\hline
    &  &  $r$ & $r_{k\leq10}$& $P$ & $P_{k\leq10}$ & $D_{Inter}$ & $D_{Inner}$ \\
\hline \multirow{5}{11 pt}{\emph{RYM}}

&PBS  & 0.063  & 0.479  & 0.036   & 0.00003  & 0.890  & 0.851   \\
 \cline{2-8}
&  HHP &     0.046  & 0.334 & 0.042 & 0.00008   &     0.936  & 0.863 \\
 \cline{2-8}
 &  OHHP &     0.046  & 0.204 & 0.040 & 0.00016   &     0.929  & 0.870 \\
 \cline{2-8}
 & DCB  & \textbf{0.046}  & \textbf{0.181}  & 0.040  & \textbf{0.00020}& \textbf{0.946}&\textbf{0.882}\\
 \hline \multirow{5}{11 pt}{\emph{Netflix}}
&PBS  & 0.049   & 0.472   & 0.055   & 0.000   & 0.435   &  0.636  \\
 \cline{2-8}
  & HHP &     0.044  & 0.428  & 0.062  &     0.00004 & 0.595  & 0.672\\
 \cline{2-8}
 & OHHP &     0.043  & 0.345  & 0.059  &     0.00009 & 0.575  & 0.686\\
\cline{2-8}
 & DCB & 0.046 & \textbf{0.343}  & 0.059  &     \textbf{0.00011}  & \textbf{0.784}  & \textbf{0.807}
\\
 \hline \multirow{5}{11 pt}{\emph{MovieLens}}
&PBS  &  0.106   & 0.573   & 0.075   & 0.000   & 0.618   & 0.645   \\
 \cline{2-8}
 &  HHP &     0.085 & 0.427 & 0.087  &     0.00020& 0.836  & 0.712 \\
 \cline{2-8}
 &  OHHP &     0.085 & 0.385 & 0.085  &     0.00037& 0.813  & 0.703 \\
 \cline{2-8}
 &DCB & 0.091 & \textbf{0.345}  & 0.081  & \textbf{0.00057} & \textbf{0.883}  & \textbf{0.764} \\
\hline
\end{tabular}
\end{center}
\end{table*}

\section{RESULTS}

To test the efficiency of the DCB algorithm, we compare the performance of the DCB with the PBS, the HHP, and the OHHP. The PBS is a pure method with a high accuracy, and the HHP is a hybrid method which resolves the dilemma between the recommendation accuracy and the diversity, and the OHHP is an improved hybrid method which further resolves the cold start problem. A summary of the performance of the PBS, the HHP, the OHHP and the DCB is presented in Tbl. 1.

\begin{table*}
\caption{The percentage improvement of the PBS, the HHP, and OHHP against the DCB in the overall ranking score $r$,
the item-degree dependent ranking score
$r_{k\leq10}$, the overall precision $P$, the item-degree dependent precision
$P_{k\leq10}$, the inter-diversity $D_{inter}$, and the inner-diversity $D_{inner}$ are shown for the \emph{RYM},
the \emph{Netflix}, and the \emph{MovieLens}, with $L=50$. To guide the eyes, if the indicator of the DCB outperforms the other methods, we show the value of the improvement percentage as a positive value, otherwise, as a negative value.}
\begin{center}
\begin{tabular}{cccccccccc}
\hline
& & $r$ & $r_{k\leq10}$& $P$ & $P_{k\leq10}$ & $D_{Inter}$ & $D_{Inner}$ \\
\hline \multirow{1}{11 pt}{\emph{RYM}} & $\delta_{PBS}$ & \textbf{37.0\%}   & \textbf{164.6\%}  & \textbf{10.0\%}  & \textbf{85.0\%}   & \textbf{5.9\%}  &\textbf{3.5\%} \\
\cline{2-8}& $\delta_{HHP}$ & 0.0\%   & \textbf{84.5\%}  & -5.0\%  & \textbf{60.0\%}   & \textbf{1.1\%}  &\textbf{2.2\%} \\
\cline{2-8}& $\delta_{OHHP}$ & 0.0\%   & \textbf{12.7\%}  & 0.0\%  & \textbf{20.0\%}   & \textbf{1.8\%}  &\textbf{1.4\%} \\
\hline \multirow{1}{11 pt}{\emph{Netflix}}  & $\delta_{PBS}$ & \textbf{6.5\%}   & \textbf{37.6\%}  & \textbf{6.8\%}  & \textbf{100.0\%}   & \textbf{44.5\%}  &\textbf{21.2\%} \\
\cline{2-8}& $\delta_{HHP}$ & -4.4\%  & \textbf{24.8\%} & -5.1\% & \textbf{63.6\%} & \textbf{24.1\%} & \textbf{16.7\%} \\
\cline{2-8}& $\delta_{OHHP}$ & -6.5\%  & \textbf{0.6\%} & 0.0\% & \textbf{18.2\%} & \textbf{26.7\%} & \textbf{15.0\%} \\
\hline \multirow{1}{11 pt}{\emph{MovieLens}} & $\delta_{PBS}$ & \textbf{16.5\%}   & \textbf{66.1\%}  & \textbf{7.4\%}  & \textbf{100.0\%}   & \textbf{30.0\%}  &\textbf{15.6\%} \\
\cline{2-8}& $\delta_{HHP}$ & -6.6\%  & \textbf{23.8\%} & -7.4\% & \textbf{64.9\%} & \textbf{5.3\%} & \textbf{6.8\%} \\
\cline{2-8}& $\delta_{OHHP}$ & -6.6\%  & \textbf{11.6\%} & -4.9\% & \textbf{35.1\%} & \textbf{7.9\%} & \textbf{8.0\%} \\
\hline
\end{tabular}
\end{center}
\end{table*}

To detect how much the DCB outperforms the other three algorithms, we define a percentage improvement $\delta_{ALG}$ by,

\begin{equation}
\delta_{ALG}= (Q_{ALG}-Q_{DCB})/Q_{DCB}, \label{eq.10}
\end{equation}

where the subhead $ALG$ refers to the investigated algorithm, and the $Q_{ALG}$ is the value of the metrics, i.e., the value of the $r$,
$r_{k\leq10}$, $P$, $P_{k\leq10}$, $D_{Inter}$ and $D_{Inner}$. The percentage improvements $\delta_{ALG}$ of the PBS, the HHP, and the OHHP against the DCB are summarized in Tbl. 2.

From Tbl. 1 and Tbl. 2, for all the three datasets, the DCB shows a great advantage in the recommendation accuracy of the low-degree objects, as well as the inter-diversity and the inner-diversity, when simultaneously keeps a high recommendation accuracy.

For the recommendation accuracy, we focus on the overall recommendation accuracy and the recommendation accuracy of the cold items. Compared with the highly accurate PBS method, the DCB outperforms the PBS for all the metrics. Taken the \emph{RYM} as an example, the DCB outperforms the PBS as much as $164.6\%$, $85.0\%$ for the recommendation accuracy of the low-degree objects $r_{k\leq10}$, $P_{k\leq10}$, $37.0\%$, $10.0\%$ for the overall recommendation accuracy $r$ and $P$, and $5.9\%$, $3.5\%$ for the inter-diversity $D_{inter}$ and the inner-diversity $D_{inner}$. Similar outstanding performance of the DCB against the PBS is also observed for the \emph{Netflix} and the \emph{MovieLens}. It indicates the DCB is highly accurate.

The HHP is an excellent algorithm in both the accuracy and the diversity, at the optimal hybridization parameter. Compared with the HHP at the optimal hybridization parameter defined by the ranking score, the DCB presents a close or very little lower overall recommendation accuracy, but an apparent great advantage in the recommendation accuracy of the cold items. For the \emph{RYM}, the DCB outperforms the HHP as much as $84.5\%$ and $60.0\%$ for the recommendation accuracy of the low-degree items $r_{k\leq10}$, $P_{k\leq10}$, but shows a close value of $0.0\%$ and $-5.0\%$ in the overall recommendation accuracy $r$ and $P$. For the \emph{Netflix} and the \emph{MovieLens}, the DCB shows a little loss of the overall recommendation accuracy $r$ and $P$, with $-4.4\%$, $-5.1\%$ for the \emph{Netflix}, and $-6.6\%$, $-7.4\%$ for the \emph{MovieLens}. However, the improvement percentage of the $r_{k\leq10}$ and the $P_{k\leq10}$ is $24.8\%$, $63.6\%$ for the \emph{Netflix}, and $23.8\%$, $64.9\%$ for the \emph{MovieLens}. The improvement percentage of the recommendation accuracy for the cold items is much higher than the loss in the overall recommendation accuracy. It further suggests that the DCB is outstanding in the cold start problem, while keeping a high recommendation accuracy. Moreover, we find that the DCB outperforms the HHP in both the inter-diversity $D_{inter}$ and the inner-diversity $D_{inner}$ for all the three datasets. The improvement percentage of the inter-diversity $D_{inter}$ and the inner-diversity $D_{inner}$ is computed to be $1.1\%$, $2.2\%$ for the RYM, and $24.1\%$, $16.7\%$ for the Netflix, and $5.3\%$, $6.8\%$ for the MovieLens.

The OHHP method is apparently advantageous in the cold start problem. Compared with the OHHP at the optimal hybridization parameter defined by the ranking score, the DCB method further greatly improves the recommendation accuracy of the cold items. For the \emph{RYM}, the DCB outperforms the OHHP as much as $12.7\%$, $20.0\%$ for the recommendation accuracy of the low-degree objects $r_{k\leq10}$, $P_{k\leq10}$, but shows a very near recommendation accuracy of the overall items. Again, similar behavior is found  for the \emph{Netflix} and the  \emph{MovieLens}.

To further understand the recommendation efficiency on the cold items, we show the degree distribution $p(k)$ of the items in the top $L=50$ recommendation list in Fig. 3. It is observed that the probability of the cold items decreases as the order of the DCB, the OHHP, the HHP, and the PBS, which indicates that the DCB indeed greatly contributes to the recommendation efficiency of the cold items.

\begin{figure}[htb]
\centering
\includegraphics[width=6cm]{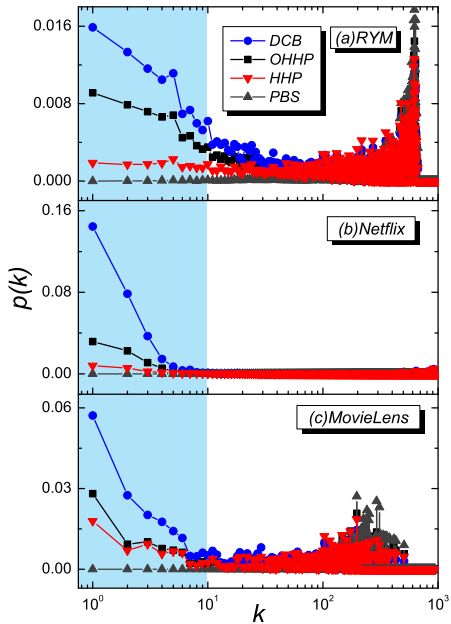}
\caption{The degree distribution $p(k)$ of the items in the top $L=50$ recommendation list is displayed. \label{Fig:3} }
\end{figure}

\begin{figure}[htb]
\centering
\includegraphics[width=6cm]{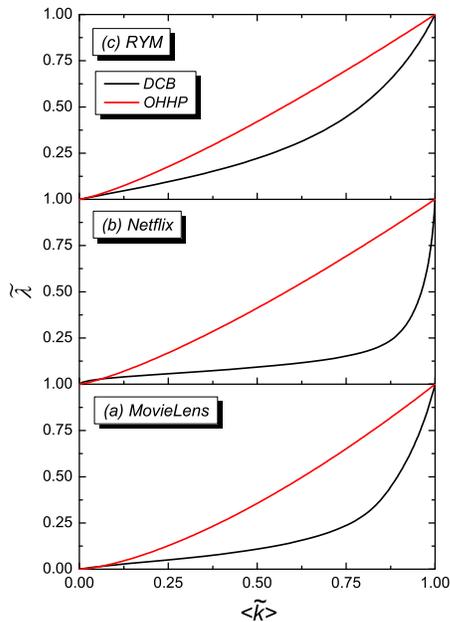}
\caption{The rescaled hybridization parameter $\widetilde{\lambda}$ on the rescaled average degree $\langle \widetilde{k} \rangle$ is displayed for the DCB and the OHHP. \label{Fig:4} }
\end{figure}

Intuitively, the improvement of the recommendation accuracy of the cold items might corresponds to the improvement of the recommendation diversity. However, by the comparison between the OHHP and the original HHP, we find that the inter-diversity of the OHHP is a little lower than that of the HHP for all the three datasets, and the inner-diversity of the \emph{RYM} and the \emph{Netflix} is a little higher than the HHP, but of the \emph{MovieLens} is also lower than that of the HHP. It suggests that the OHHP does not show apparent advantages in the recommendation diversity, though it greatly improves the recommendation accuracy of the cold items.

To better understand the observed phenomena, we show the rescaled hybridization parameter $\widetilde{\lambda}$ on the rescaled average degree $\langle \widetilde{k} \rangle$ of the OHHP and the DCB in Fig. 4, where the curve of the DCB is obtained from the empirical study. It is observed that the $\widetilde{\lambda}$ on the $\langle \widetilde{k} \rangle$ of the OHHP deviates the empirical curve of the DCB, which can partly explain why the OHHP method unilaterally improves the recommendation accuracy of the cold-items, but not simultaneously enhances the recommendation diversity. Compared with the OHHP, the DCB not only further improves the recommendation accuracy of the cold items, but also elevates the recommendation diversity.

Further investigation of the inter-diversity $D_{inter}$ on the recommendation list length $L$ suggests that, for all the four methods, the inter-diversity decreases with the recommendation list length $L$, as shown in Fig. 5. It is reasonable since the difference between different users recommendation list will decreases with the augment of the recommendation list length $L$. Compared with the pure method of the PBS, the hybrid methods of the HHP and the OHHP show a much slower decay of the inter-diversity. Especially, the DCB exhibits a much higher value and a much slower decay of the inter-diversity for the overall range of the recommendation list length $L$. It indicates that the recommendation diversity of the DCB is not only higher but also more stable than the PBS, the HHP, and the OHHP, with the recommendation list length.

The inner-diversity $D_{inner}$ on the recommendation list length $L$ is shown in Fig. 6. For the \emph{RYM}, it is observed that the $D_{inner}$ increases with $L$ for all the four algorithms. However, for the \emph{Netflix} and the \emph{MovieLens}, the $D_{inner}$ increases with $L$ only for the PBS, the HHP, the OHHP, but exhibits a very stable and high value for the DCB. It further suggests that the DCB provides a highly and steadily diverse recommendation.

\begin{figure}[htb]
\centering
\includegraphics[width=6cm]{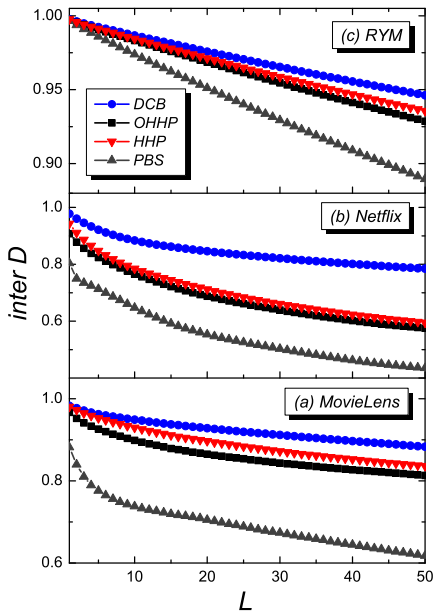}
\caption{The inter-diversity $D_{inter}$ on the recommendation list length $L$ is displayed.\label{Fig:5} }
\end{figure}

\begin{figure}[htb]
\centering
\includegraphics[width=6cm]{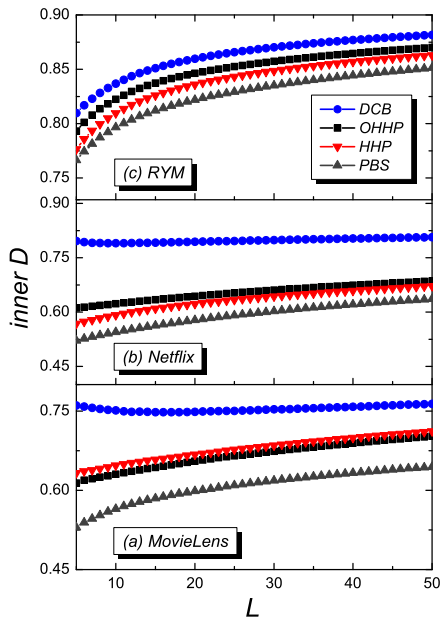}
\caption{The inner-diversity $D_{inner}$ on the recommendation list length $L$ is displayed. \label{Fig:6} }
\end{figure}

Taken together, while not requiring any procedure of the optimal hybridization parameter selection according to a specific evaluator, but adaptively assigning the optimal parameter according to the relational function between the algorithm and the item degree, the DCB remarkably outperforms the PBS, the HHP, and the OHHP in the recommendation accuracy of cold items, as well as the recommendation diversity, and simultaneously keeps a high overall recommendation accuracy.

The dilemma existing most in common in hybrid algorithms is how to choose proper optimal hybridization parameter according to different recommendation focuses. It is out of question that recommendation accuracy is one of the most important evaluators of the algorithm performance. However, even when evaluating recommendation accuracy, different indicators might take different values. By relating the data property to the algorithm, we resolve the explicit dilemma of the hybridization parameter selection for the complex contradiction among different recommendation focuses.

Moreover, the cold start problem is a long-standing challenging in traditional recommendation system, since it is difficult for users to be aware of the cold items with insufficient accessorial information\cite{rus08,lee08}. However, for most systems, the cold items occupy a big proportion. In the \emph{RYM}, the \emph{Netflix}, and the \emph{Movielens}, the cold items whose degrees are no more than 10 are as much as $24.56 \%$, $50.62 \%$, and $41.26 \%$. Developing efficient information filtering techniques is essentially required to solve the cold start problem. Integrating the tag information has been taken as an efficient way to make prediction for cold items \cite{zha10b}, which however increases the system complexity. The DCB greatly improves the recommendation accuracy of the cold items, whereas keeps a high overall accuracy, from the perspective of constructing the possible correlation between the algorithm design and the dataset characteristic.

Furthermore, most past studies overwhelmingly emphasize the recommendation accuracy, but underestimate the importance of diversity. In fact, diversity can well evaluate the personal recommendation. However, recommendation accuracy and diversity are an apparent dilemma pair in traditional information filtering system. Typical examples are the PBS and the HTS algorithms, where the PBS is more accurate but less diverse, whereas the HTS is more diverse but less accurate. The DCB method shows an excellent recommendation diversity, as well as a high recommendation accuracy, by finding out a recommendation list length independent relational function between the hybridization parameter and the item degrees.

\section{CONCLUSION}

In this article, we propose a data characteristic based recommendation algorithm by finding out the relational function between the hybridization parameter and the item degrees. We use a combined exponential function form $\widetilde{\lambda}=ae^{b \langle \widetilde{k}
\rangle}+ce^{d\langle \widetilde{k} \rangle}$ to fit the relation curve, which is independent of the recommendation list length. With this implementation, hybridization parameters are adaptively obtained according to the specific individual item degree. Experimental results show that, the proposed method significantly promotes the performance of the long-standing cold start problem, as well as the recommendation diversity, while simultaneously keeps a high recommendation accuracy, without requiring any additional accessory information.

Previous studies show that, most pure methods perform excellently in either recommendation accuracy or diversity, whereas hybrid methods generally outperform in both accuracy and diversity at an optimal hybridization parameter. However, how to obtain the real optimal hybridization parameter for different recommendation focuses still remains controversial. In this article, we have shown that, the dilemma of seeking for the optimal parameter of hybrid methods can be elegantly resolved by constructing the relational function between the algorithm and the dataset characteristic.

Furthermore, the cold start problem is a long-standing challenge in traditional recommendation systems. Due to very little accessorial information of the cold items, it is hard for users to be aware of these items. Utilizing tag information has been taken as an efficient way to solve the cold start problem, which however increases the system complexity. The manifested DCB method shows a great promotional effect on the cold item recommendation accuracy, as well as the diversity. Our present work might shed some new light on the personal recommendation.

\begin{acknowledgments}
This work was partially supported by the National Natural Science Foundation
of China (Grant Nos. 11175079, 10805025, 11105024, 61103109 and 60973069).
\end{acknowledgments}

\bibliography{recomm}

\begin{thebibliography}{34}
\expandafter\ifx\csname natexlab\endcsname\relax\def\natexlab#1{#1}\fi
\expandafter\ifx\csname bibnamefont\endcsname\relax
  \def\bibnamefont#1{#1}\fi
\expandafter\ifx\csname bibfnamefont\endcsname\relax
  \def\bibfnamefont#1{#1}\fi
\expandafter\ifx\csname citenamefont\endcsname\relax
  \def\citenamefont#1{#1}\fi
\expandafter\ifx\csname url\endcsname\relax
  \def\url#1{\texttt{#1}}\fi
\expandafter\ifx\csname urlprefix\endcsname\relax\def\urlprefix{URL }\fi
\providecommand{\bibinfo}[2]{#2}
\providecommand{\eprint}[2][]{\url{#2}}

\bibitem[{\citenamefont{Masum and Zhang}(2004)}]{mas04}
\bibinfo{author}{\bibfnamefont{H.}~\bibnamefont{Masum}} \bibnamefont{and}
  \bibinfo{author}{\bibfnamefont{Y.-C.} \bibnamefont{Zhang}},
  \bibinfo{journal}{First Monday} \textbf{\bibinfo{volume}{9}},
  \bibinfo{pages}{7} (\bibinfo{year}{2004}).

\bibitem[{\citenamefont{Adomavicius and Tuzhilin}(2005)}]{ado05}
\bibinfo{author}{\bibfnamefont{G.}~\bibnamefont{Adomavicius}} \bibnamefont{and}
  \bibinfo{author}{\bibfnamefont{A.}~\bibnamefont{Tuzhilin}},
  \bibinfo{journal}{IEEE Trans. Knowl.Data Eng.} \textbf{\bibinfo{volume}{17}},
  \bibinfo{pages}{734} (\bibinfo{year}{2005}).

\bibitem[{\citenamefont{Goldberg et~al.}(1992)\citenamefont{Goldberg, Nichols,
  Oki, and Terry}}]{gol92}
\bibinfo{author}{\bibfnamefont{D.}~\bibnamefont{Goldberg}},
  \bibinfo{author}{\bibfnamefont{D.}~\bibnamefont{Nichols}},
  \bibinfo{author}{\bibfnamefont{B.~M.} \bibnamefont{Oki}}, \bibnamefont{and}
  \bibinfo{author}{\bibfnamefont{D.}~\bibnamefont{Terry}},
  \bibinfo{journal}{Commun. ACM} \textbf{\bibinfo{volume}{35}},
  \bibinfo{pages}{61} (\bibinfo{year}{1992}).

\bibitem[{\citenamefont{Schafer et~al.}(2007)\citenamefont{Schafer, Frankowski,
  Herlocker, and Sen}}]{sch07}
\bibinfo{author}{\bibfnamefont{J.~B.} \bibnamefont{Schafer}},
  \bibinfo{author}{\bibfnamefont{D.}~\bibnamefont{Frankowski}},
  \bibinfo{author}{\bibfnamefont{J.}~\bibnamefont{Herlocker}},
  \bibnamefont{and} \bibinfo{author}{\bibfnamefont{S.}~\bibnamefont{Sen}},
  \bibinfo{journal}{Lect. Notes. Comput. Sc.} \textbf{\bibinfo{volume}{4321}},
  \bibinfo{pages}{291} (\bibinfo{year}{2007}).

\bibitem[{\citenamefont{Pazzani and Billsus}(2007)}]{paz07}
\bibinfo{author}{\bibfnamefont{M.~J.} \bibnamefont{Pazzani}} \bibnamefont{and}
  \bibinfo{author}{\bibfnamefont{D.}~\bibnamefont{Billsus}},
  \bibinfo{journal}{Lect. Notes. Comput. Sc.} \textbf{\bibinfo{volume}{4321}},
  \bibinfo{pages}{325} (\bibinfo{year}{2007}).

\bibitem[{\citenamefont{Balabanovic and Shoham}(1997)}]{bal97}
\bibinfo{author}{\bibfnamefont{M.}~\bibnamefont{Balabanovic}} \bibnamefont{and}
  \bibinfo{author}{\bibfnamefont{Y.}~\bibnamefont{Shoham}},
  \bibinfo{journal}{Comm. ACM} \textbf{\bibinfo{volume}{40}},
  \bibinfo{pages}{66} (\bibinfo{year}{1997}).

\bibitem[{\citenamefont{Goldberg et~al.}(2001)\citenamefont{Goldberg, Roeder,
  Gupta, and Perkins}}]{gol01}
\bibinfo{author}{\bibfnamefont{K.}~\bibnamefont{Goldberg}},
  \bibinfo{author}{\bibfnamefont{T.}~\bibnamefont{Roeder}},
  \bibinfo{author}{\bibfnamefont{D.}~\bibnamefont{Gupta}}, \bibnamefont{and}
  \bibinfo{author}{\bibfnamefont{C.}~\bibnamefont{Perkins}},
  \bibinfo{journal}{Infor. Retr.} \textbf{\bibinfo{volume}{4}},
  \bibinfo{pages}{133} (\bibinfo{year}{2001}).

\bibitem[{\citenamefont{Maslov and Zhang}(2001)}]{mas01}
\bibinfo{author}{\bibfnamefont{S.}~\bibnamefont{Maslov}} \bibnamefont{and}
  \bibinfo{author}{\bibfnamefont{Y.-C.} \bibnamefont{Zhang}},
  \bibinfo{journal}{Phys. Rev. Lett.} \textbf{\bibinfo{volume}{87}},
  \bibinfo{pages}{248701} (\bibinfo{year}{2001}).

\bibitem[{\citenamefont{Hofmann}(2004)}]{hof04}
\bibinfo{author}{\bibfnamefont{T.}~\bibnamefont{Hofmann}},
  \bibinfo{journal}{ACM. Trans. Inf. Syst.} \textbf{\bibinfo{volume}{22}},
  \bibinfo{pages}{89} (\bibinfo{year}{2004}).

\bibitem[{\citenamefont{Blei et~al.}(2003)\citenamefont{Blei, Ng, and
  Jordan}}]{ble03}
\bibinfo{author}{\bibfnamefont{D.~M.} \bibnamefont{Blei}},
  \bibinfo{author}{\bibfnamefont{A.~Y.} \bibnamefont{Ng}}, \bibnamefont{and}
  \bibinfo{author}{\bibfnamefont{M.~I.} \bibnamefont{Jordan}},
  \bibinfo{journal}{J. Mach. Learn. Res.} \textbf{\bibinfo{volume}{3}},
  \bibinfo{pages}{993} (\bibinfo{year}{2003}).

\bibitem[{\citenamefont{Laureti et~al.}(2006)\citenamefont{Laureti, Moret,
  Zhang, and Yu}}]{lau06}
\bibinfo{author}{\bibfnamefont{P.}~\bibnamefont{Laureti}},
  \bibinfo{author}{\bibfnamefont{L.}~\bibnamefont{Moret}},
  \bibinfo{author}{\bibfnamefont{Y.-C.} \bibnamefont{Zhang}}, \bibnamefont{and}
  \bibinfo{author}{\bibfnamefont{Y.~K.} \bibnamefont{Yu}},
  \bibinfo{journal}{EPL} \textbf{\bibinfo{volume}{75}}, \bibinfo{pages}{1006}
  (\bibinfo{year}{2006}).

\bibitem[{\citenamefont{Ren et~al.}(2008)\citenamefont{Ren, Zhou, and
  Zhang}}]{ren08}
\bibinfo{author}{\bibfnamefont{J.}~\bibnamefont{Ren}},
  \bibinfo{author}{\bibfnamefont{T.}~\bibnamefont{Zhou}}, \bibnamefont{and}
  \bibinfo{author}{\bibfnamefont{Y.-C.} \bibnamefont{Zhang}},
  \bibinfo{journal}{EPL} \textbf{\bibinfo{volume}{82}}, \bibinfo{pages}{58007}
  (\bibinfo{year}{2008}).

\bibitem[{\citenamefont{Zhang et~al.}(2011)\citenamefont{Zhang, Zhou, and
  Y.-C.}}]{zha11}
\bibinfo{author}{\bibfnamefont{Z.-K.} \bibnamefont{Zhang}},
  \bibinfo{author}{\bibfnamefont{T.}~\bibnamefont{Zhou}}, \bibnamefont{and}
  \bibinfo{author}{\bibfnamefont{Z.}~\bibnamefont{Y.-C.}}, \bibinfo{journal}{J.
  Comput. Sci. Technol.} \textbf{\bibinfo{volume}{26}}, \bibinfo{pages}{767}
  (\bibinfo{year}{2011}).

\bibitem[{\citenamefont{Zhang et~al.}(2007)\citenamefont{Zhang, Blattner, and
  Yu}}]{zha07}
\bibinfo{author}{\bibfnamefont{Y.-C.} \bibnamefont{Zhang}},
  \bibinfo{author}{\bibfnamefont{M.}~\bibnamefont{Blattner}}, \bibnamefont{and}
  \bibinfo{author}{\bibfnamefont{Y.~K.} \bibnamefont{Yu}},
  \bibinfo{journal}{Phys. Rev. Lett.} \textbf{\bibinfo{volume}{99}},
  \bibinfo{pages}{154301} (\bibinfo{year}{2007}).

\bibitem[{\citenamefont{Zhou et~al.}(2007)\citenamefont{Zhou, Ren, Medo, and
  Zhang}}]{zho07}
\bibinfo{author}{\bibfnamefont{T.}~\bibnamefont{Zhou}},
  \bibinfo{author}{\bibfnamefont{J.}~\bibnamefont{Ren}},
  \bibinfo{author}{\bibfnamefont{M.}~\bibnamefont{Medo}}, \bibnamefont{and}
  \bibinfo{author}{\bibfnamefont{Y.-C.} \bibnamefont{Zhang}},
  \bibinfo{journal}{Phys. Rev. E} \textbf{\bibinfo{volume}{76}},
  \bibinfo{pages}{046115} (\bibinfo{year}{2007}).

\bibitem[{\citenamefont{Jia et~al.}(2008)\citenamefont{Jia, Liu, Sun, and
  Wang}}]{jia08}
\bibinfo{author}{\bibfnamefont{C.-X.} \bibnamefont{Jia}},
  \bibinfo{author}{\bibfnamefont{R.-R.} \bibnamefont{Liu}},
  \bibinfo{author}{\bibfnamefont{D.}~\bibnamefont{Sun}}, \bibnamefont{and}
  \bibinfo{author}{\bibfnamefont{B.-H.} \bibnamefont{Wang}},
  \bibinfo{journal}{Physica A} \textbf{\bibinfo{volume}{387}},
  \bibinfo{pages}{5887} (\bibinfo{year}{2008}).

\bibitem[{\citenamefont{Zhou et~al.}(2009)\citenamefont{Zhou, Su, Liu, Jiang,
  Wang, and Zhang}}]{zho09}
\bibinfo{author}{\bibfnamefont{T.}~\bibnamefont{Zhou}},
  \bibinfo{author}{\bibfnamefont{R.~Q.} \bibnamefont{Su}},
  \bibinfo{author}{\bibfnamefont{R.-R.} \bibnamefont{Liu}},
  \bibinfo{author}{\bibfnamefont{L.~L.} \bibnamefont{Jiang}},
  \bibinfo{author}{\bibfnamefont{B.-H.} \bibnamefont{Wang}}, \bibnamefont{and}
  \bibinfo{author}{\bibfnamefont{Y.-C.} \bibnamefont{Zhang}},
  \bibinfo{journal}{New J. Phys.} \textbf{\bibinfo{volume}{11}},
  \bibinfo{pages}{123008} (\bibinfo{year}{2009}).

\bibitem[{\citenamefont{Liu and Deng}(2009)}]{liu09a}
\bibinfo{author}{\bibfnamefont{J.}~\bibnamefont{Liu}} \bibnamefont{and}
  \bibinfo{author}{\bibfnamefont{G.~S.} \bibnamefont{Deng}},
  \bibinfo{journal}{Physica A} \textbf{\bibinfo{volume}{388}},
  \bibinfo{pages}{3643} (\bibinfo{year}{2009}).

\bibitem[{\citenamefont{Zhou et~al.}(2010)\citenamefont{Zhou, Kuscsik, Liu,
  Medo, Wakeling, and Zhang}}]{zho10}
\bibinfo{author}{\bibfnamefont{T.}~\bibnamefont{Zhou}},
  \bibinfo{author}{\bibfnamefont{Z.}~\bibnamefont{Kuscsik}},
  \bibinfo{author}{\bibfnamefont{J.-G.} \bibnamefont{Liu}},
  \bibinfo{author}{\bibfnamefont{M.}~\bibnamefont{Medo}},
  \bibinfo{author}{\bibfnamefont{J.~R.} \bibnamefont{Wakeling}},
  \bibnamefont{and} \bibinfo{author}{\bibfnamefont{Y.-C.} \bibnamefont{Zhang}},
  \bibinfo{journal}{Proc. Natl. Acad. Sci. U.S.A.}
  \textbf{\bibinfo{volume}{107}}, \bibinfo{pages}{4511} (\bibinfo{year}{2010}).

\bibitem[{\citenamefont{Liu et~al.}(2009{\natexlab{a}})\citenamefont{Liu, Jia,
  Zhou, Sun, and Wang}}]{liu09b}
\bibinfo{author}{\bibfnamefont{R.-R.} \bibnamefont{Liu}},
  \bibinfo{author}{\bibfnamefont{C.-X.} \bibnamefont{Jia}},
  \bibinfo{author}{\bibfnamefont{T.}~\bibnamefont{Zhou}},
  \bibinfo{author}{\bibfnamefont{D.}~\bibnamefont{Sun}}, \bibnamefont{and}
  \bibinfo{author}{\bibfnamefont{B.-H.} \bibnamefont{Wang}},
  \bibinfo{journal}{Physica A} \textbf{\bibinfo{volume}{388}},
  \bibinfo{pages}{462} (\bibinfo{year}{2009}{\natexlab{a}}).

\bibitem[{\citenamefont{Marchette and Priebe}(2008)}]{mar08}
\bibinfo{author}{\bibfnamefont{D.~J.} \bibnamefont{Marchette}}
  \bibnamefont{and} \bibinfo{author}{\bibfnamefont{C.~E.}
  \bibnamefont{Priebe}}, \bibinfo{journal}{Comput. Statist. Data Anal.}
  \textbf{\bibinfo{volume}{52}}, \bibinfo{pages}{1373} (\bibinfo{year}{2008}).

\bibitem[{\citenamefont{Zhang et~al.}(2010{\natexlab{a}})\citenamefont{Zhang,
  Zhou, and Zhang}}]{zha10a}
\bibinfo{author}{\bibfnamefont{Z.-K.} \bibnamefont{Zhang}},
  \bibinfo{author}{\bibfnamefont{T.}~\bibnamefont{Zhou}}, \bibnamefont{and}
  \bibinfo{author}{\bibfnamefont{Y.-C.} \bibnamefont{Zhang}},
  \bibinfo{journal}{Physica A} \textbf{\bibinfo{volume}{389}},
  \bibinfo{pages}{179} (\bibinfo{year}{2010}{\natexlab{a}}).

\bibitem[{\citenamefont{Shang et~al.}(2010)\citenamefont{Shang, Zhang, Zhou,
  and Zhang}}]{sha10}
\bibinfo{author}{\bibfnamefont{M.-S.} \bibnamefont{Shang}},
  \bibinfo{author}{\bibfnamefont{Z.-K.} \bibnamefont{Zhang}},
  \bibinfo{author}{\bibfnamefont{T.}~\bibnamefont{Zhou}}, \bibnamefont{and}
  \bibinfo{author}{\bibfnamefont{Y.-C.} \bibnamefont{Zhang}},
  \bibinfo{journal}{Physica A} \textbf{\bibinfo{volume}{389}},
  \bibinfo{pages}{1259} (\bibinfo{year}{2010}).

\bibitem[{\citenamefont{Liu et~al.}(2009{\natexlab{b}})\citenamefont{Liu, Zhou,
  Wang, and Zhang}}]{liu09c}
\bibinfo{author}{\bibfnamefont{J.-G.} \bibnamefont{Liu}},
  \bibinfo{author}{\bibfnamefont{T.}~\bibnamefont{Zhou}},
  \bibinfo{author}{\bibfnamefont{B.-H.} \bibnamefont{Wang}}, \bibnamefont{and}
  \bibinfo{author}{\bibfnamefont{Y.-C.} \bibnamefont{Zhang}},
  \bibinfo{journal}{Int. J. Mod. Phys. C} \textbf{\bibinfo{volume}{20}},
  \bibinfo{pages}{1925} (\bibinfo{year}{2009}{\natexlab{b}}).

\bibitem[{\citenamefont{Liu et~al.}(2010)\citenamefont{Liu, Zhou, Che, Wang,
  and Zhang}}]{liu10a}
\bibinfo{author}{\bibfnamefont{J.-G.} \bibnamefont{Liu}},
  \bibinfo{author}{\bibfnamefont{T.}~\bibnamefont{Zhou}},
  \bibinfo{author}{\bibfnamefont{H.~A.} \bibnamefont{Che}},
  \bibinfo{author}{\bibfnamefont{B.-H.} \bibnamefont{Wang}}, \bibnamefont{and}
  \bibinfo{author}{\bibfnamefont{Y.-C.} \bibnamefont{Zhang}},
  \bibinfo{journal}{Physica A} \textbf{\bibinfo{volume}{389}},
  \bibinfo{pages}{881} (\bibinfo{year}{2010}).

\bibitem[{\citenamefont{L\"{u} and Liu}(2011)}]{lv11}
\bibinfo{author}{\bibfnamefont{L.}~\bibnamefont{L\"{u}}} \bibnamefont{and}
  \bibinfo{author}{\bibfnamefont{W.}~\bibnamefont{Liu}},
  \bibinfo{journal}{Phys. Rev. E} \textbf{\bibinfo{volume}{83}},
  \bibinfo{pages}{066119} (\bibinfo{year}{2011}).

\bibitem[{\citenamefont{Liu et~al.}(2011)\citenamefont{Liu, Zhou, and
  Guo}}]{liu11}
\bibinfo{author}{\bibfnamefont{J.-G.} \bibnamefont{Liu}},
  \bibinfo{author}{\bibfnamefont{T.}~\bibnamefont{Zhou}}, \bibnamefont{and}
  \bibinfo{author}{\bibfnamefont{Q.}~\bibnamefont{Guo}},
  \bibinfo{journal}{Phys. Rev. E} \textbf{\bibinfo{volume}{84}},
  \bibinfo{pages}{037101} (\bibinfo{year}{2011}).

\bibitem[{\citenamefont{Zhang et~al.}(2010{\natexlab{b}})\citenamefont{Zhang,
  Liu, Zhang, and Zhou}}]{zha10b}
\bibinfo{author}{\bibfnamefont{Z.-K.} \bibnamefont{Zhang}},
  \bibinfo{author}{\bibfnamefont{C.}~\bibnamefont{Liu}},
  \bibinfo{author}{\bibfnamefont{Y.-C.} \bibnamefont{Zhang}}, \bibnamefont{and}
  \bibinfo{author}{\bibfnamefont{T.}~\bibnamefont{Zhou}},
  \bibinfo{journal}{EPL} \textbf{\bibinfo{volume}{92}}, \bibinfo{pages}{28002}
  (\bibinfo{year}{2010}{\natexlab{b}}).

\bibitem[{\citenamefont{Ahn}(2008)}]{ahn08}
\bibinfo{author}{\bibfnamefont{H.~J.} \bibnamefont{Ahn}},
  \bibinfo{journal}{Inf. Sci.} \textbf{\bibinfo{volume}{178}},
  \bibinfo{pages}{37} (\bibinfo{year}{2008}).

\bibitem[{\citenamefont{Herlocker et~al.}(2004)\citenamefont{Herlocker,
  Konstan, Terveen, and Riedl}}]{her04}
\bibinfo{author}{\bibfnamefont{J.~L.} \bibnamefont{Herlocker}},
  \bibinfo{author}{\bibfnamefont{J.~A.} \bibnamefont{Konstan}},
  \bibinfo{author}{\bibfnamefont{L.~G.} \bibnamefont{Terveen}},
  \bibnamefont{and} \bibinfo{author}{\bibfnamefont{J.~T.} \bibnamefont{Riedl}},
  \bibinfo{journal}{ACM Trans. Inf. Syst.} \textbf{\bibinfo{volume}{22}},
  \bibinfo{pages}{5} (\bibinfo{year}{2004}).

\bibitem[{\citenamefont{Qiu et~al.}(2011)\citenamefont{Qiu, Chen, Zhang, and
  Zhou}}]{qiu11a}
\bibinfo{author}{\bibfnamefont{T.}~\bibnamefont{Qiu}},
  \bibinfo{author}{\bibfnamefont{G.}~\bibnamefont{Chen}},
  \bibinfo{author}{\bibfnamefont{Z.-K.} \bibnamefont{Zhang}}, \bibnamefont{and}
  \bibinfo{author}{\bibfnamefont{T.}~\bibnamefont{Zhou}},
  \bibinfo{journal}{EPL} \textbf{\bibinfo{volume}{95}}, \bibinfo{pages}{58003}
  (\bibinfo{year}{2011}).

\bibitem[{\citenamefont{Zhou et~al.}(2008)\citenamefont{Zhou, Jiang, Su, and
  Zhang}}]{zho08}
\bibinfo{author}{\bibfnamefont{T.}~\bibnamefont{Zhou}},
  \bibinfo{author}{\bibfnamefont{L.~L.} \bibnamefont{Jiang}},
  \bibinfo{author}{\bibfnamefont{R.~Q.} \bibnamefont{Su}}, \bibnamefont{and}
  \bibinfo{author}{\bibfnamefont{Y.-C.} \bibnamefont{Zhang}},
  \bibinfo{journal}{EPL} \textbf{\bibinfo{volume}{81}}, \bibinfo{pages}{58004}
  (\bibinfo{year}{2008}).

\bibitem[{\citenamefont{Russell and Yoon}(2008)}]{rus08}
\bibinfo{author}{\bibfnamefont{S.}~\bibnamefont{Russell}} \bibnamefont{and}
  \bibinfo{author}{\bibfnamefont{V.}~\bibnamefont{Yoon}},
  \bibinfo{journal}{Expert Syst. Appl.} \textbf{\bibinfo{volume}{34}},
  \bibinfo{pages}{2316} (\bibinfo{year}{2008}).

\bibitem[{\citenamefont{Lee et~al.}(2008)\citenamefont{Lee, Park, and
  Park}}]{lee08}
\bibinfo{author}{\bibfnamefont{T.~Q.} \bibnamefont{Lee}},
  \bibinfo{author}{\bibfnamefont{Y.}~\bibnamefont{Park}}, \bibnamefont{and}
  \bibinfo{author}{\bibfnamefont{Y.~T.} \bibnamefont{Park}},
  \bibinfo{journal}{Expert Syst. Appl.} \textbf{\bibinfo{volume}{34}},
  \bibinfo{pages}{3055} (\bibinfo{year}{2008}).

\end{thebibliography}

\end{document}